# CALCULATING VORONOI DIAGRAMS USING SIMPLE CHEMICAL REACTIONS

BEN DE LACY COSTELLO

*Unconventional Computing Centre,*
*University of the West of England, Bristol, UK, BS161QY*
*E-mail Ben.DeLacyCostello@uwe.ac.uk*

ABSTRACT

This paper overviews work on the use of simple chemical reactions to calculate Voronoi diagrams and undertake other related geometric calculations. This work highlights that this type of specialised chemical processor is a model example of a parallel processor. For example increasing the complexity of the input data within a given area does not increase the computation time. These processors are also able to calculate two or more Voronoi diagrams in parallel. Due to the specific chemical reactions involved and the relative strength of reaction with the substrate (and cross-reactivity with the products) these processors are also capable of calculating Voronoi diagrams sequentially from distinct chemical inputs.
The chemical processors are capable of calculating a range of generalised Voronoi diagrams (either from circular drops of chemical or other geometric shapes made from adsorbent substrates soaked in reagent) , skeletonisation of planar shapes and weighted Voronoi diagrams (e.g. additively weighted Voronoi diagrams, Multiplicatavely weighted Crystal growth Voronoi diagrams). The paper will also discuss some limitations of these processors.
These chemical processors constitute a class of pattern forming reactions which have parallels with those observed in natural systems. It is possible that specialised chemical processors of this general type could be useful for synthesising functional structured materials.

## 1. Introduction

Voronoi diagrams are used to partition space into spheres of influence in many branches of science and can be observed in numerous natural systems. They find uses in many fields such as astronomy (galaxy cluster analysis[1]), biology (modelling tumour cell growth [2]), chemistry (modelling crystal growth [3]), ecology (modelling competition [4]) and computational geometry (solving nearest neighbour problems [5]). In addition to standard and generalised Voronoi diagrams weighted diagrams are commonly used for example to model crystal growth [6]. In simple terms an Additively weighted Voronoi diagram (AWVD) is where all sources grow at the same rate but some start at different times. A Multiplicatively weighted Voronoi diagram (MWVD) is where all sources start growing at the same time but some grow with different rates. The multiplicatively weighted crystal growth Voronoi diagram (MWCGVD) [6] is an extension of the MWVD and overcomes the problem of disconnected regions which are not possible in growing crystals (or expanding chemical fronts). Therefore, in the MWCGVD when regions with large growth rates meet regions with lower growth rates they can "wrap around" these regions to form a connected Voronoi diagram of the plane. The bisectors of weighted Voronoi diagrams are hyperbolic segments rather than straight lines. For a more in depth description and mathematical treatment of Voronoi diagrams and weighted derivatives see references [6,7]

In addition to their many uses in modeling, Voronoi diagrams can be observed in nature, for example, in animal coat markings [8] between interacting bacterial, fungal or slime mould colonies [9,10] between growing crystals and even in universal structures such as gravitational caustics [11]. They were also observed in Gas discharge systems [12]. The formation of Voronoi diagrams in chemical systems with one reagent and one substrate was first reported in [13]. This and similar reactions were utilised as chemical processors for the computation of the shortest obstacle free path [14], skeltonisation of a planar shape [15,16] and construction of a prototype XOR gate [17]. The formation of weighted Voronoi diagrams from two reagents reacted on one substrate (potassium ferrocyanide gel) were first reported in [18]. In the same year the calculation of three Voronoi diagrams in parallel was reported in a chemical system based on the reaction of two reagents on a mixed substrate gel (potassium ferrocyanide and potassium ferricyanide) [19]. More recently complex tessellations of the plane have been reported where one of two binary reagents exhibited limited or no reactivity with the gel substrate [20]. It was shown that re-useable processors for calculation of Voronoi diagrams could be constructed using a crystallisation process [21]. However, these processors must be reset completely prior to carrying out sequential calculations. Subsequent to this it was found that by carefully selecting reagents it was possible to calculate two or more Voronoi diagrams sequentially on the same substrate. Sequential calculations are useful in comparing overlapping Voronoi regions.

A new class of pattern forming inorganic reaction based on the reaction of sodium hydroxide with copper chloride loaded gels was reported [22]. These reactions exhibited a range of self-organised patterns such as cardioid and spiral waves only observed previously in more complex chemical reactions such as the BZ reaction. This is highly significant given the very simple nature of the chemical reactants.

More recently another inorganic system has been identified based on aluminium chloride gels reacted with sodium hydroxide [23]. This system is particularly remarkable as unlike the system mentioned above [22] the wave evolution can be easily observed in real-time.

In previous work we were able to exert control over another class of inorganic reaction with the same primary wave splitting mechanism [24]. The reaction involved the addition of copper chloride to potassium ferricyanide immobilised in an agarose gel. At high concentrations of potassium ferricyanide circular (cone shaped) waves are spontaneously generated by heterogeneities and expand. Where these waves collide they form a natural tessellation of the plane similar to those obtained in crystallisation processes. By reducing the concentration of potassium ferricyanide then the point and time at which circular waves are intiated could be controlled by marking the gel with a very fine glass needle.

By using this methodology user defined precipitation patterns could be created that equate to Voronoi tessellations of the plane. In previous work [25] involving the aluminium chloride sodium hydroxide reaction we adopted the same strategy and were able to show that initiation of travelling waves was possible. However, marking the gel prior to initiation of the reaction produced complex travelling waves made up of multiple spiral wave fragments. In further work [26] it was established that marking the gel after initiation (pouring on the outer electrolyte) allowed selection of cardioid or circular waves, depending on the time after initiation. Therefore, the construction of Voronoi diagrams and additively weighted Voronoi diagrams in this system proved possible.

Section 2 of the paper relates to the construction of Voronoi diagrams and related geometric calculations in stable chemical systems. Section 3. Relates to the construction of Voronoi diagrams in unstable chemical systems. Section 4. Compares the systems from Section 2 and 3 with other unconventional parallel processors, both chemical and biological. Section 5. Gives some general conclusions

**2.1 Construction of Voronoi Diagrams in Stable Systems**

*2.2 Simple Voronoi Diagram Construction*

2.1.1 Potassium ferricyanide or ferrocyanide loaded gels

For the purposes of this paper we will deal mainly with reactions based on the reaction of potassium ferrocyanide or ferricyanide loaded gels reacted with various metal salts. However, a huge variety of inorganic reactions studied display this phenomena in the right concentration ranges (usually low substrate concentration and high outer electrolyte concentration). To undertake the reactions agar Gel (Sigma, St Louis, USA, 0.3g) was added to deionised water (30ml) and heated with stirring to 70$^o$C and then removed from the heat. Potassium ferricyanide (Fisher Scientific, Leicestershire, UK, 75mg (7.59mM)) or Potassium ferrocyanide (BDH chemicals Ltd, Poole, UK, 75mg (5.91mM)) was added with stirring. The solution was then poured into five 9cm diameter Petri dishes and left to set for one hour. All possible binary combinations of the metal ions listed below were reacted on both potassium ferrocyanide and ferricyanide gels. The metal ions used in construction of the Voronoi tessellations were as follows: Iron (III) nitrate [300mg/ml, 1.24M] (can be substituted with chloride salt), Iron (II) sulphate [300mg/ml, 1.97M], silver (III) nitrate [300mg/ml, 1.76M], cobalt (II) chloride hexahydrate [300mg/ml, 1.26 M], lead (II) chloride [300mg/ml, 1.07M], manganese (II) chloride tetrahydrate [300mg/ml, 1.51M], chromium (III) nitrate nonahydrate [300mg/ml, 1.26M], nickel (II) chloride [300mg/ml, 2.31M] and copper(II) chloride [300mg/ml, 2.23M]. To form simple Voronoi diagrams drops of these metal ion solutions are carefully placed on the surface of the gel. Figure 1a shows a Voronoi diagram constructed when ferric chloride was reacted with potassium ferrocyanide. The drops of ferric chloride were placed in three different regular patterns to highlight the versatility of this approach. Fig 1b shows a simple Voronoi diagram calculated in the reaction between cupric chloride and potassium ferrocyanide. It should be noted that in these chemical systems bisector width is proportional to the distance between initiation sources.

2.1.2. Palladium chloride loaded gels

For the purposes of these experiments a reaction–diffusion processor based on palladium chloride was used. A gel of agar (2% by weight) containing palladium chloride (in the range 0.2–0.9% by weight, for the images displayed the lower range of 0.2% was favoured) was prepared by mixing the solids in warm deionised water.

The mixture was heated with a naked flame until it boiled with constant stirring to ensure full dissolution of palladium chloride and production of a uniform gel (on cooling). The liquid was then rapidly transferred to petri dishes. The outer electrolyte used was a saturated solution (at 20$^o$C) of potassium iodide. Figure 1c shows a generalised Voronoi diagram constructed in the palladium chloride potassium iodide chemical processor.

*2.2. Generalised Voronoi Diagram Construction*

Actually all Voronoi diagram construction using chemical inputs in stable systems are generalised as the input is not a point source. However, circular drops of reagent construct an almost identical Voronoi diagram to that expected for a point source. In this section generalised refers to an input which is not circular.

It was found that filter paper soaked in the outer electrolyte could be used to initiate reaction fronts rather than drops of the outer electrolyte. Therefore, generalized Voronoi diagrams could be constructed i.e. where the sources of wave generation are geometric shapes rather than point sources. In this case the bisectors are curved rather than straight lines. This kind of computation is not trivial for conventional computer processors. However, for the parallel chemical processors the complexity of the computational task is almost irrelevant. Figure 2 shows an example of a generalized Voronoi diagram formed on a potassium ferrocyanide gel.

*2.3. Skeletonisation of a Planar Shape*

This reaction can be used to calculate the internal Voronoi diagram (skeleton) of a contour. This is a useful data reduction method for image recognition etc. Figure 3. Shows the skeletonisation of a geometric contour. The skeleton of a pentagon as calculated by the chemical reaction is a five pointed star. This is because as mentioned the distance between the opposed initiation fronts controls bisector width.

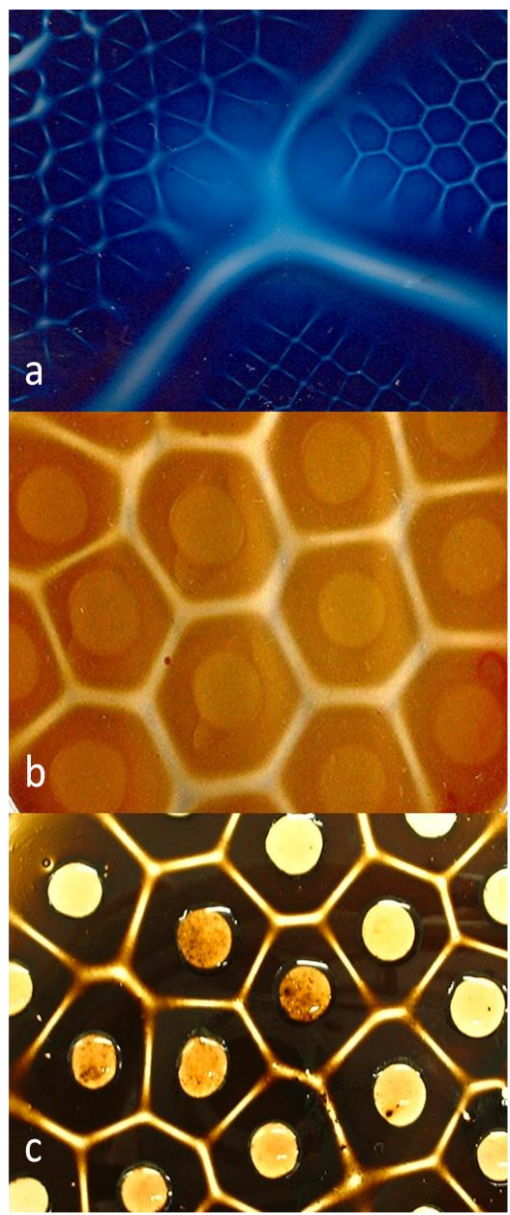

Figure 1. Voronoi diagrams formed in stable chemical systems via the application of drops of reagents. 1a. Voronoi diagram formed when ferric chloride was reacted with a potassium ferrocynaide loaded gel. The reactant drops (ferric chloride) were placed in various array patterns in different sections of the petri dish. The bisectors with low precipitate concentration are clearly visible. 1b. Voronoi diagram formed when copper chloride was reacted with a potassium ferrocyanide loaded gel. In this case the position of the original drops and the bisector formation can be observed. 1c. Voronoi diagram formation when potassium iodide was reacted with a palladium chloride loaded gel. In this case the original positions of the drops and the bisectors appear as low precipitate regions, probably due to redissolution of the primary product in the outer electrolyte (in contrast to the ferricyanide/ferrocyanide based systems).

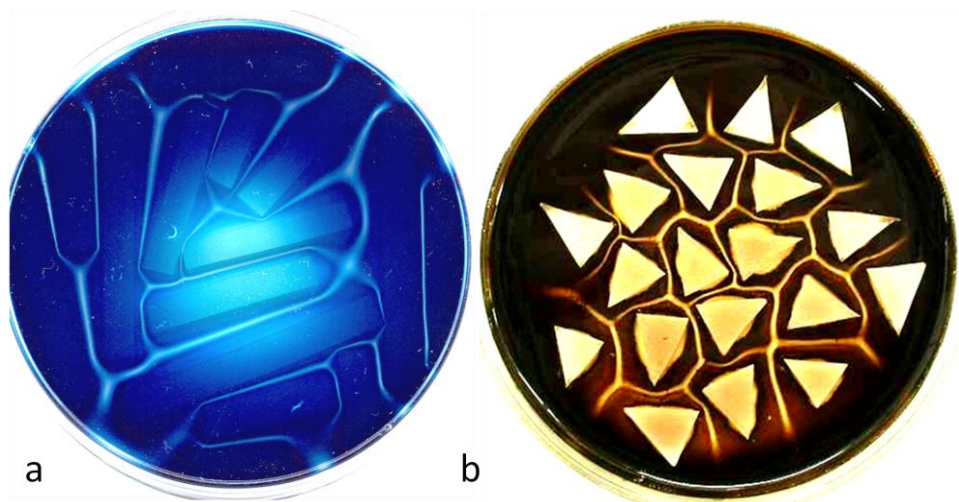

Figure 2. Generalised Voronoi diagrams constructed from geometric shapes (other than circles). 2a. Generalised Voronoi diagram formed when ferric chloride soaked filter paper was reacted with a potassium ferrocyanide loaded gel. 2b. Generalised Voronoi diagram when potassium ioide soaked filter paper was reacted with a palladium chloride loaded gel.

*2.4 Mixed Cell Voronoi Diagrams*

2.4.1 Balanced binary Voronoi tessellations
Two or more different metal ion solutions can be reacted with the substrate loaded gel in parallel to create a series of colourful tessellations of the plane. If the reagents have the same diffusion rate and react fully with the substrate gel then a simple Voronoi diagram is constructed (albeit that the cells are different colours). Figure 4 shows an example of a mixed cell Voronoi diagram of this type. If the reagents had equal reactivity with the gel but differing diffusion rates then a multiplicatively weighted crystal growth Voronoi diagram (MWCGVD) would be constructed. Figure 5 shows a MWCGVD constructed in this type of reaction. The bisectors are slightly curved between the cells formed via the reaction of cupric ions and the cells formed via the reaction of ferric ions indicating that the diffusion rate of the cupric ions is slightly faster.

2.4.2 Unbalanced binary Voronoi tessellations
Figure 6 shows an example of the tessellation formed when a partially reactive metal ion solution (ferric chloride) is reacted on a gel alongside a fully reactive metal ion solution (cobaltous chloride). In this reaction fronts are initiated from drops of both reactant solutions but when they meet the fronts of the fully reactive reactant cross the fronts formed by the partially reactive reagent and only annihilate (in a precipitate free region) when they meet fronts emanating from another fully reactive source. Therefore, the resulting tessellation is a Voronoi diagram of all initiation sites and a Voronoi diagram of only sites containing the fully reactive reagent. So two Voronoi diagrams can be calculated in parallel using these simple chemical reactions.

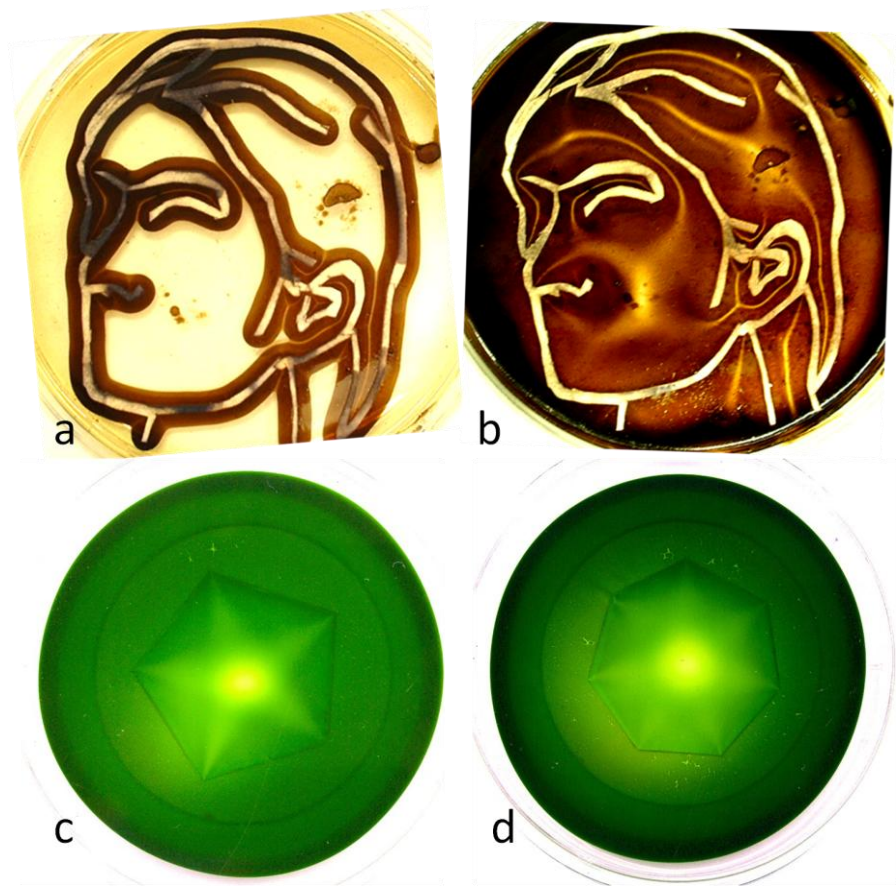

Figure 3. Skeletonisation of planar shapes or contours using a similar technique to that used to form generalised voronoi diagrams in Figure 2. Figure 3a and b shows a contour soaked in potassium iodide reacted on a palladium chloride gel. The bisectors represent a data reduction of the original contour. As the chemical reactions give bisectors which vary in width according to the distance separating the reaction fronts of the contour then additional information is encoded in the computed skeleton. This is apparent in figure 3c and d which show skeletons formed when ferrous sulphate soaked filter paper was reacted with a potassium ferricyanide loaded gel. In this case the skeleton of a pentagon is a five pointed star and of a heptagon a 7 pointed star. It is apparent that the chemical processor more successfully constructs the skeleton of the pentagon. This is because eventually a geometric shape with more sides would approximate a circle in which case the skeleton would be a circular region of reduced precipitate. Thus increasing the angle between two sides and reducing the length of individual features affects the accuracy of the calculation. This is because chemical fronts do not maintain the shape of the original boundary over long distances but adopt a distance dependent curvature.

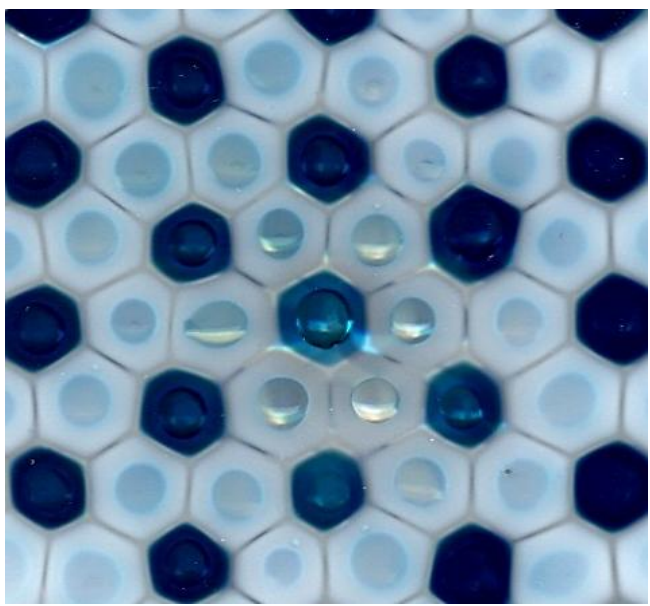

Figure 4. Mixed cell Voronoi diagram formed when ferrous sulphate and manganese chloride were reacted with a potassium ferrocyanide loaded gel.

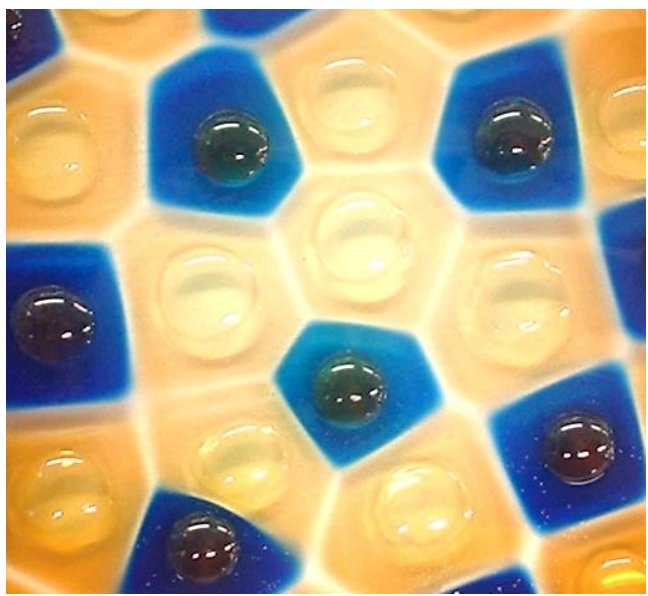

Figure 5. Multiplicatively weighted Crystal Growth Voronoi diagram formed when cupric chloride and ferric chloride were reacted on potassium ferrocyanide gel.

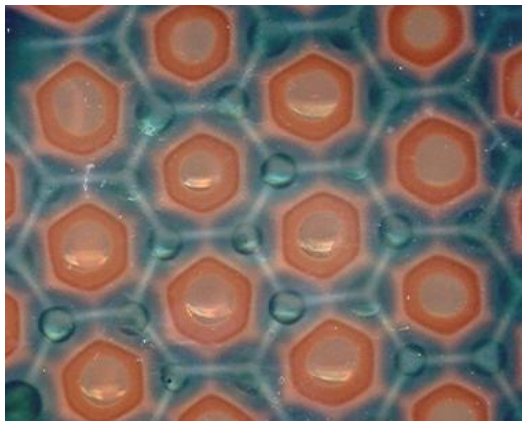

Figure 6. Complex tessellation created via the reaction of ferric chloride and cobaltous chloride on potassium ferricyanide gel.

2.4.3 Cross reactive binary Voronoi tessellations

Figure 7 shows a fully cross reactive tessellation in progress. In this case both the primary reactants produce coloured cells and a simple Voronoi diagram. However, where the fronts meet both primary products are cross reactive and a secondary precipitate is formed where the fronts overlap. Figure 8. Shows the completed reaction where only a white precipitate remains across most of the petri dish. However, what is remarkable is that the precipitate has a highly complex and ordered tessellation made up of high and low precipitate regions. It is possible that this type of reaction may be useful in directed materials synthesis if the mechanism can be better understood.

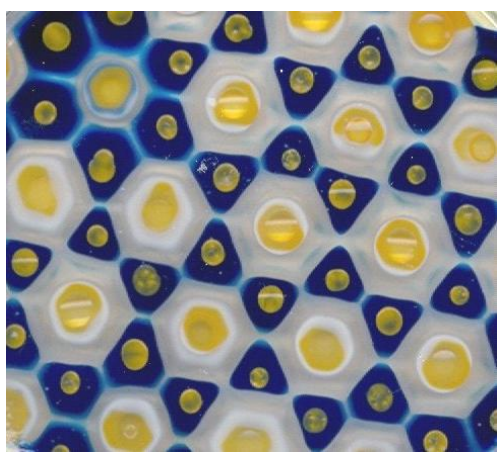

Figure 7. A Tessellation formed when two reagents with cross reactive products (ferrous sulphate and silver nitrate) are reacted on a potassium ferricyanide gel. The reaction is still in progress.

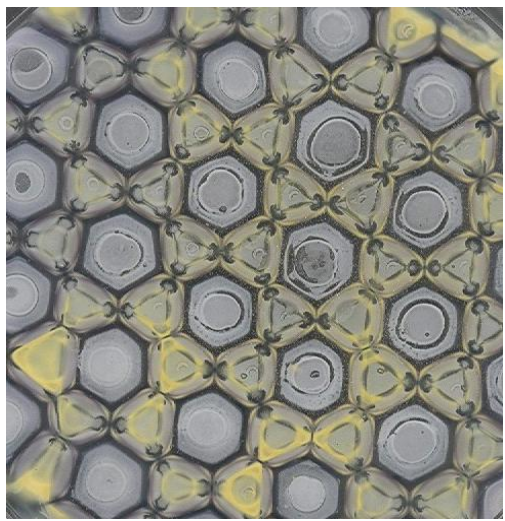

Figure 8. A Tessellation formed when two reagents with cross reactive products (ferrous sulphate and silver nitrate) are reacted on a potassium ferricyanide gel. Completed reaction.

2.4.4 Balanced and unbalanced tertiary Voronoi diagrams

The number of distinct reagents reacted on the same substrate gel can be increased to three or more. Unlike the binary case where the limit of possibilities has been fully mapped, tertiary Voronoi tessellations have only been partially classified [27]. However, they fit into the same general types, balanced (all reagents are reactive equally with the gel), unbalanced (one or more reagent is partially reactive) and cross reactive. These tertiary tessellations could be computing MWCGVD depending on the relative diffusion and reaction rates of the primary reagents. What is certain is that they produce an amazing array of colourful tessellations of the plane. Any design can be recreated provided that it can be subdivided into symmetrical regions which are spaced evenly (and allowing for partial and cross-reactivity, or incorporating them into the design). Figure 9 shows examples of this type of tessellation.

*2.5 Tessellations formed by Combining Two Exclusive Chemical Couples*

Ferric ions do not react with potassium ferricyanide and ferrous ions do not react with potassium ferrocyanide. By using a mixed gel of potassium ferricyanide and ferrocyanide and reacting this with drops of ferric ions and ferrous ions a complex tessellation can be obtained see Figure 10. This tessellation equates to three separate Voronoi diagrams constructed in parallel. One pertaining to all reactant drops and two which are exclusive to each set of different reactant drops. These are calculated where both sets of fronts cross and only annihilate in a precipitate free region where they meet another front from their exclusive chemical couple.

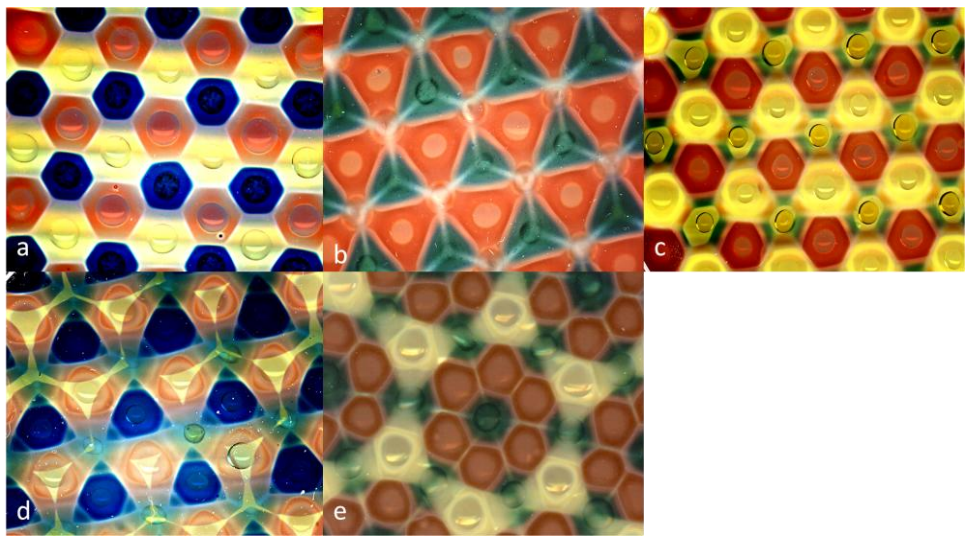

Figure 9 Showing a range of tertiary tessellations (3 metal ion reagents reacted simultaneously on the substrate gel) of the plane. Figure 9a showing a balanced tessellation when ferrous sulphate (blue), nickel (II) chloride (yellow) and cobalt (II) chloride (red) was reacted on potassium ferricyanide gel Figure 9b showing a tessellation formed where cobalt (II) chloride (red), iron (III) nitrate (partially reactive) and chromium (III) nitrate (unreactive were reacted on potassium ferricyanide gel. Figure 9c showing tessellation when ferric chloride (partially reactive), nickel (II) chloride and cobalt chloride were reacted on potassium ferricyanide gel. Figure 9b showing a cross reactive formed when iron (III) nitrate, copper (II) chloride and iron (II) sulphate were reacted on potassium ferricyanide gel 9e tessellation where cobalt chloride, nickel chloride and ferric chloride were reacted on potassium ferricyanide gel.

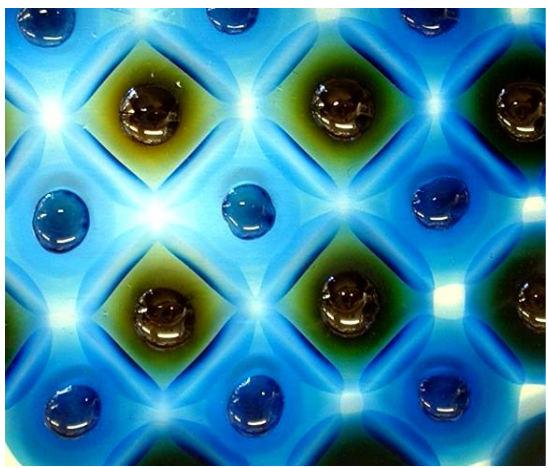

Figure 10. Tessellation formed when ferrous ions and ferric ions are reacted on a mixed potassium ferricyanide and potassium ferricyanide gel.

*2.6 Sequential Voronoi Diagram Calculation*

Previously it had been assumed that these chemical processors were single use in terms of sequential parallel geometric computations. However, the careful selection of primary and secondary reactants based on their reactivity with the substrate gel, allows for sequential calculations to be undertaken (see Figure 11). Thus a reagent with partial reactivity with the substrate is first reacted to give a Voronoi diagram at which point subsequent points are added using a reactant with full reactivity (alternatively use of reagents whose products have mutual cross reactivity can be undertaken to achieve a similar result).

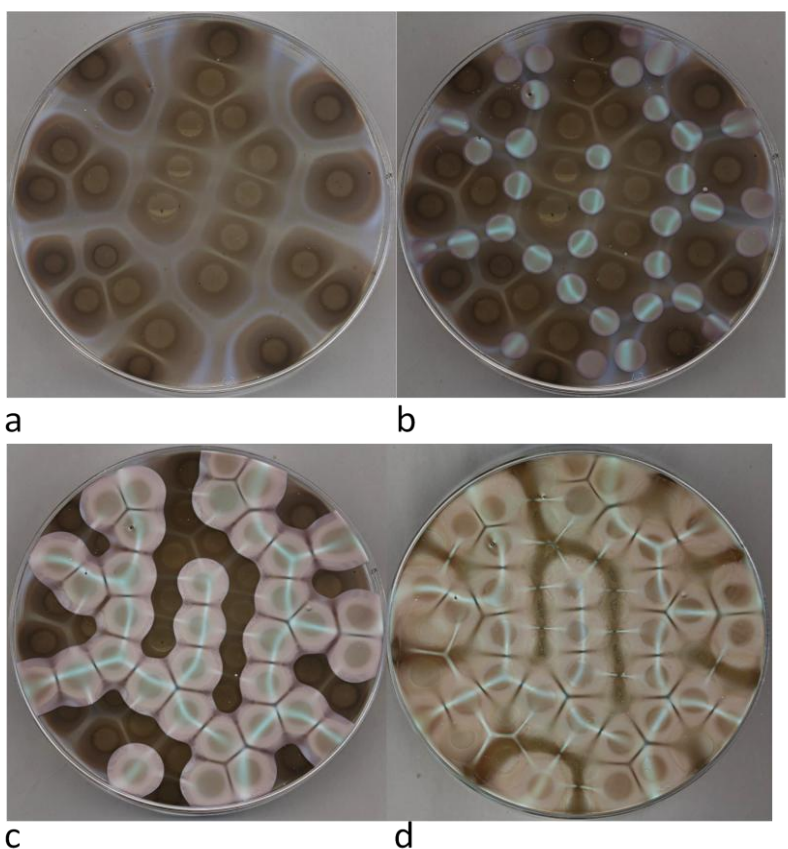

Figure 11a. Voronoi diagram formed by $Ag^+$ ions reacted on potassium ferrocyanide gel. b. Addition of Secondary $Co^{2+}$ reactant to the gel ubstrate containing the original Voronoi diagram. It can be seen that a physical precipitation reaction is intiated. c. 45 minutes into the reaction, $Co^{2+}$ diffusing fronts interact to form secondary Voronoi diagram bisectors (precipitate free regions). d. Final reaction showing formation of distinctive interacting Voronoi diagrams. The bisectors formed in these chemical reactions inherently code for distance. Therefore, points/sites separated by larger distances produce wide bisectors. This is useful when used in optimal path planning and minimal data reduction via the skeletonisation of planar shapes.

*2.7 Speed of Computation*

In these stable reactions the speed of Voronoi diagram calculation is relatively slow at the scales we have presented. The reason for conducting these experiments in this manner was to ensure a visual output to the computation, and ease of adding the reagents in parallel to the substrate. However, as mentioned the complexity of adding three or more reagents or calculating upto 3 Voronoi diagrams in parallel has no real affect on the computation time within a defined area. Furthermore, if the reagent density is increased and the input size decreased then the computation is much faster (of the order of a few seconds). This could presumably be reduced further by optimisation of the gel system and identifying a method such as inket printing to add nanolitre sized droplets in parallel to the substrate surface. Figure 12. shows a Voronoi diagram computed using a manually applied high drop density. The drop volume is calculated to be in the nanolitre range.

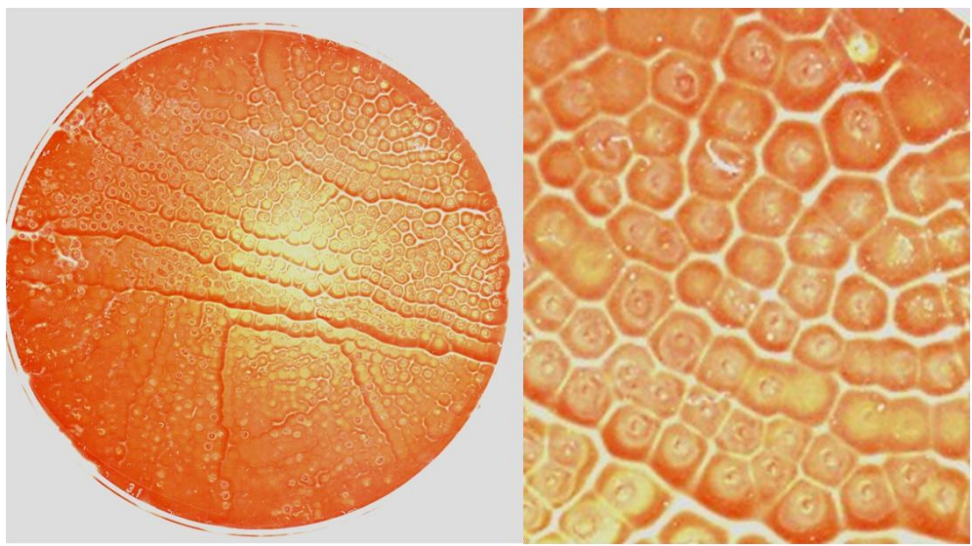

Figure 12. Voronoi diagram calculated from hundreds of nanolitre sized droplets of copper chloride reacted on a potassium ferrocyanide gel. The computation was complete in under 30 seconds. This compares to a time of an hour or more for larger drops spaced according to the tessellations previously presented in the paper.

**3. Formation of Voronoi diagrams in unstable systems**

*3.1 Failure to Compute a Voronoi diagram*

Figure 13. shows the results of raising the potassium ferrocyanide concentration from 2.5mg/ml (5.91mM) to 7.5mg/ml (17.73mM). At this concentration the circular fronts can be observed to spontaneously split meaning that a complete Voronoi diagram is not formed. This splitting of the reaction front is qualitatively similar to the primary splitting

mechanism described by Hantz [22] for the copper chloride sodium hydroxide reaction. In this phase of the reaction extended pattern formation can be observed. Interestingly Figure 13 shows Voronoi bisectors calculated at points where no fronts meet (due to the instability in the front, causing the progress of one or more reaction fronts to cease) but should have met if the fronts had maintained stability. This does highlight the long range forces that must be acting in advance of the fronts.

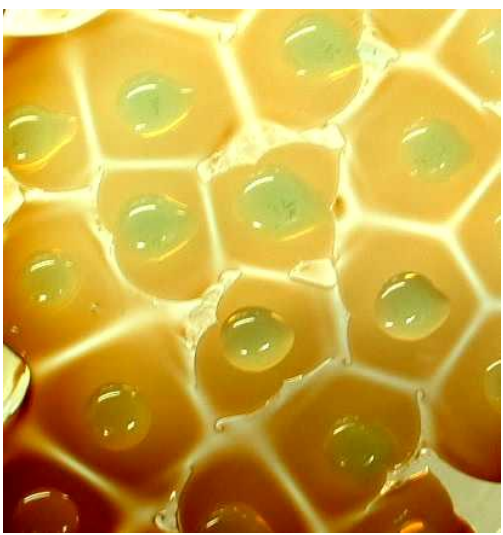

Figure 13. Failure to compute Voronoi diagram due to chemical instability which disrupts the circular fronts

*3.2 Spontaneous Voronoi Diagram Formation in Cupric Chloride Potassium Ferro/Ferricyanide Systems*

If the concentration of inner electrolyte is raised further to a concentration circa 30 mg/ml (70.92mM) then circular reaction fronts do not form but where solution has been applied to the gel a number of small expanding circular waves can be observed. Because these waves represent a splitting of the front in three dimensions they are actually expanding cone shaped regions enclosing unreacted substrate. Where these cone shaped regions collide they form tessellations which when observed in 2D equate to Voronoi diagrams of the original points of instability see Figure 14.

*3.3 Controlled Voronoi Diagram Formation in Cupric Chloride Potassium Ferro/Ferricyanide Systems*

By reducing the concentration of potassium ferrocyanide or potassium ferricyanide in the gel a reaction where spontaneous wave formation was minimised was obtained. However, when the gel was pre-marked with a thin glass needle prior to initiation of the reaction this provided a large enough heterogeneity in the system after initiation (pouring on outer electrolyte) to selectively initiate a cone shaped wave from this point. Therefore, programmable Voronoi diagrams could be constructed in these unstable systems see Figure 15. For experimental details refer to [24]. Figure 16. Shows a schematic of how Voronoi diagrams are formed from interacting precipitation fronts.

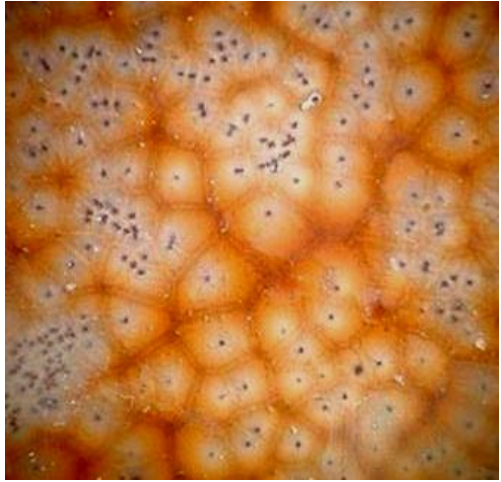

Figure 14. spontaneous "Voronoi diagram" formation when copper chloride solution 300mg/ml (2.24M) is poured over a potassium ferrocyanide gel (30mg/ml, 70.92mM).

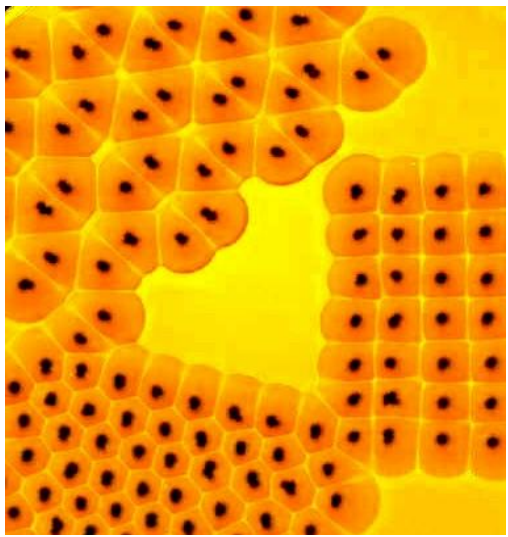

Figure 15. Controlled Voronoi diagram construction in an unstable system based on the reaction of cupric chloride (2.93M) with a potassium ferricyanide gel (0.09M). The gel was pre-marked with the pattern using a glass needle (diameter 0.5mm). Then the outer electrolyte solution of cupric chloride was poured over the gel to inititiate the 3D precipitate waves

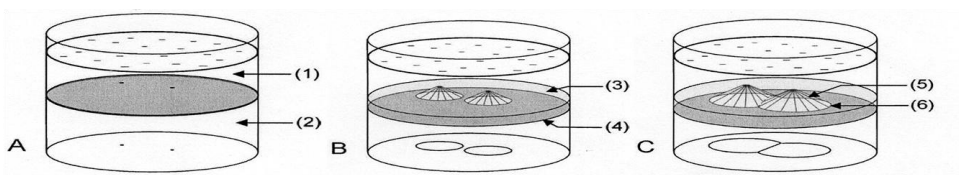

Figure 16 Schematic representation of Voronoi diagram formation through the ''regressing edge mechanism.'' 1 - Outer electrolyte, 2 - agarose gel containing the inner electrolyte, 3- precipitate, 4 -reaction front active border, 5- passive border, 6- regressing edge. The contours of the patterns are drawn on the bottom of the vessel. A – The precipitation does not start at some points of the gel surface. B - These points expand into precipitate-free cones as the front progresses. C - The regressing edges meet on the line that is an equal distance from the tip of the empty cones to form a Voronoi diagram (in a two dimensional view).

*3.4 Controlling Wave Generation in the Aluminium Chloride Sodium Hydroxide Reaction*

Lagzi and co-workers described the generation of moving self organised structures in a simple precipitation system [23]. Structures such as circular waves, cardioid and spiral waves can be observed at different concentration ranges of the reaction. For details of the experimental procedures used see [23-26]. We constructed a phase diagram for the reaction in order to identify whether there was a controllable phase in this reaction see Figure 17. By choosing concentration ranges in the controllable region but close to the spontaneous pattern generation region of the phase diagram it was possible to initiate disorganised spiral waves from specific locations by marking the gel with a fused silica needle (diameter 0.25mm) prior to initiation of the reaction see Figure 18. It did not prove possible to initiate circular waves selectively at any concentration ranges within this region using this technique. Therefore, we marked the gel at specific time intervals after the initiation of the reaction. Figure 19. Shows a petri dish where a set of three points have been marked at increasing time intervals after the reaction has been initiated. What this shows is that there is a tendency towards single circular wave generation with increasing time interval after initiation. Therefore, if the gel is marked two minutes after initiation single circular waves can be generated consistently. Interestingly, it was possible to generate cardioid waves fairly reproducibly if the gel was marked 60 seconds after initiation. In addition we observed that in this sytem the formation of single circular waves was due to the failure of an expanding wave fragment to form a cardioid wave. Thus the spiral tips were annihilated in the collision resulting in a seemingly circular expanding wave.

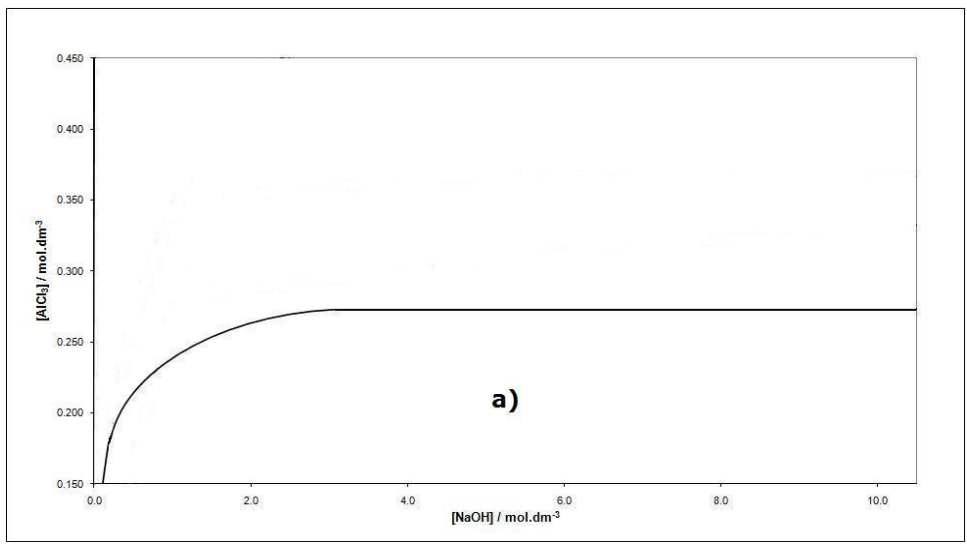

Figure 17. Phase diagram for the aluminium chloride sodium hydroxide reaction identifying the controllable region a) where waves can be generated by marking the gel prior to addition of sodium hydroxide solution. In the region outside a) we observed spontaneous wave generation.

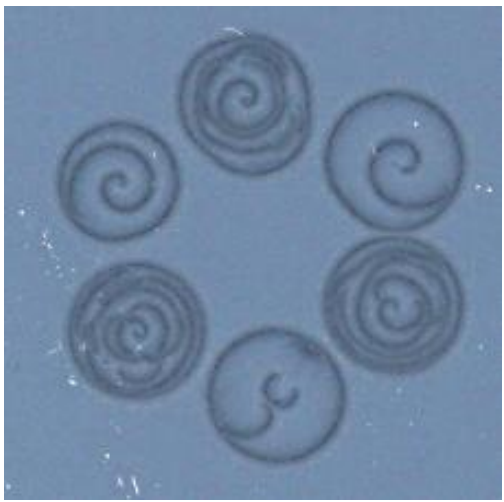

Figure 18. Controlled evolution of six complex waves in the aluminium chloride (0.26M) and sodium hydroxide reaction (5M). The gel was marked once at each of six positions in a hexagonal pattern. After marking the gel the sodium hydroxide was poured onto the gel to initiate the precipitation front.

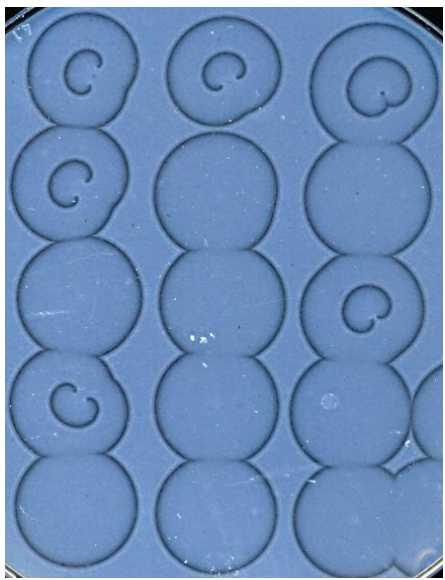

Figure 19. The selective initiation of waves in the aluminium chloride (0.26M) sodium hydroxide (5M) reactions. The gel was marked (from top to bottom) 60 seconds after initiation and subsequently at 15 second intervals with the final set of three points marked 2 minutes after initiation.

3.4.1 Selective initiation of circular waves

Therefore, by marking the gel at time intervals greater than 2 minutes circular waves rather than disorganised waves (obtained if gel was marked prior to initiation) can be reproducibly obtained see Figure 20. Re-marking the gel at the centre of the circular front results in the initiation of a second circular front see Figure 21.

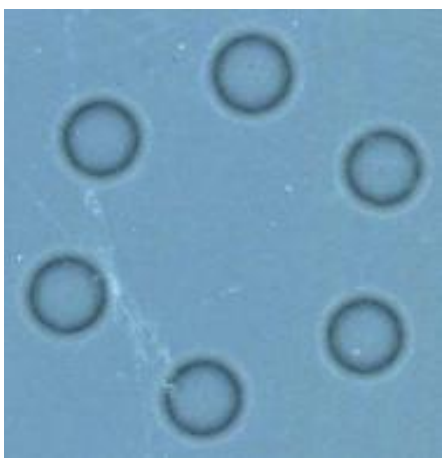

Figure 20. The selective initiation of circular waves in the aluminium chloride (0.26M) sodium hydroxide (5M) reaction. The gel was marked 5 minutes after initiation.

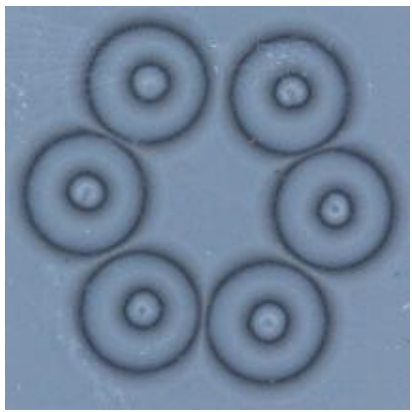

Figure 21. Initiation of nested circular waves in the aluminium chloride (0.26M) sodium hydroxide (5M) reaction.

*2.5 Construction of a Voronoi diagram*

Therefore, if circular waves can be initiated reproducibly and the circular fronts annihilate in straight line bisectors then a Voronoi diagram can be constructed see Figure 22.

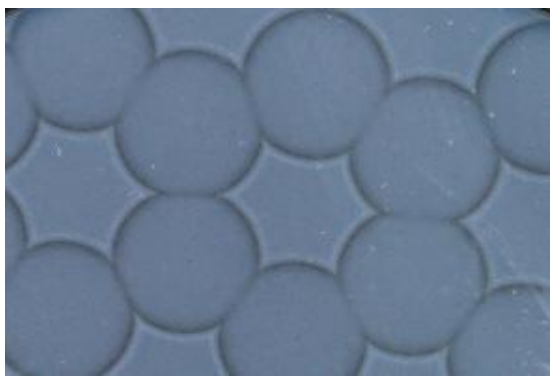

Figure 22. Construction of a Voronoi diagram in the aluminium chloride (0.26M) sodium hydroxide (5M) reaction. Reaction in progress.

2.5.1 Construction of an additively weighted Voronoi diagram
A further step is to construct weighted diagrams. In this system it is relatively easy to construct an Additively weighted Voronoi diagram by marking the gel at different locations at various time intervals after the initiation of the reaction see Figure 23. The curved bisectors between the smaller cells initiated at longer time intervals after the start of the reaction and the other larger cells signifies correct calculation of the weighted diagram.

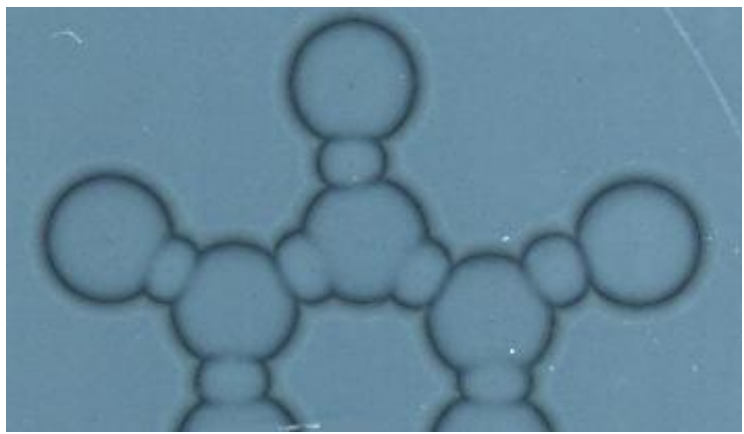

Figure 23. Additively weighted Voronoi diagram under construction in the aluminium chloride (0.26M) sodium hydroxide (5M) reaction.

## 4. Discussion

*4.1 Stable Vs Unstable Systems*

It has been demonstrated that Voronoi diagrams can be formed in both stable systems where input is in the form of drops of reagents (such as metal ions) which diffuse and react with a substrate loaded gel [13-16,18-20, 27-28] and unstable systems [24-26] either spontaneously (presumably at heterogenities in the system which trigger wave evolution) or in controlled systems (where concentration is held just below the level likely to result in natural wave evolution and large physical heterogeneities are applied to the system in order to trigger waves). It is interesting to note that stable and unstable Voronoi diagram formation can occur in two distinct phases of the same reactions (i.e. the reagents have not changed only the concentration of the reagents). It is also interesting to note that Voronoi diagram formation (generalized) is computed by both stable and unstable systems but additively weighted diagrams are not computed correctly by stable systems (studied so far) and multiplicatively weighted diagrams (crystal growth type) are not computed correctly by unstable systems (studied so far). This is in contrast to crystal growth whereby crystals within the same domain can grow with different speeds and start growing at different times relative to each other. Although unstable systems are similar in some way to crystallization processes, they rely predominantly on the diffusion rate of the outer electrolyte through the gel to control growth rate of the waves. Thus at each point in the evolution of the reaction the growth rate of the initiated waves will be almost identical.

The advantages of unstable systems vs. stable systems are:

- Voronoi diagram formation is from point source (not circular region) – accuracy of calculation increased
- Additively weighted diagram can easily be calculated - by marking the gel before and after initiation (adding the reagent). In stable systems additively weighted diagrams cannot be calculated correctly as fronts seem to exert attractive forces.
- Expanding precipitate waves exert repulsive forces leading to correct bisector calculation in AWVD
- Ease of data input. This is because inputs for calculation are added prior to initiation of the reaction.
- Potential to design functional materials in 3D
- Faster computation (potentially) and on smaller length scales

The disadvantages of unstable systems vs. stable systems are:

- Methods of multiple parallel computations yet to be established
- Programmable multiplicatively weighted Voronoi diagram cannot be computed simply - as wave expansion velocity is hard coded to the reaction (whereas in stable systems diffusion and reaction rates of different reactants which can be reacted on the same substrate naturally differ and this info can be utilised to undertake specific computations.)
- Cannot be used for sequential calculations.
- The computational output is not as obvious in some reactions and may require post image processing to extract data.

*4.2 Computational Limitations*

As mentioned in the previous sections, these chemical processors have some potential limitations from a computational perspective. These are dealt with in some detail in this previous paper [28]. The accuracy of computational output is limited by the diffusion and reaction velocity of the chemical species and the distance between sources. This is because fronts are subject to natural curvature which increases with time. This limits the accuracy of the calculated bisectors especially if the source of waves is a planar shape with acute angles (as these will not be reproduced in the growing front). In the calculation of a skeleton another problem exists, and in fact acute angles allow more accurate reconstruction than obtuse angles where the fronts lose definition and approximate to a circular wavefront. Also in the calculation of a skeleton the data reduction is limited. This is because the fronts do not just interact to form output at maximal points of interaction but continuously across the substrate resulting in a 3-dimensional data reduction. Although this is useful for reconstruction of the exact contour, data reduction is limited *per se*. In the previous paper [28] the problem of inverting a Voronoi diagram was studied and it was found that due to the inherent mechanism of the chemical construction it proved impossible to recreate the original data point set from a calculated Voronoi diagram.

There are specific problems in calculating weighted Voronoi diagrams in stable and unstable systems. As mentioned the stable systems easily calculate a MWCGVD but fail to compute an AWVD. This is because even though it is easy to add point sources which diffuse at the same speed at distinct time points in the reactions evolution, bisector construction is not correct. Figure 24 shows an example of one such calculation. This shows that the bisectors are not computed correctly according to the theoretical output. In fact the bisectors are inverted completely from those expected theoretically. This suggests that an attractive force is responsible for the mechanism of bisector formation rather than a repulsive force which would be required for correct calculation. In unstable systems it is the opposite problem calculation of an additively weighted diagram is facile, but calculation of a mutiplicatively weighted diagram is problematic because at a given point in the reaction evolution all growing precipitation fronts in the unstable system are controlled only by diffusion and reaction rate and these are uniform at a given point in the progress of the reactant front.

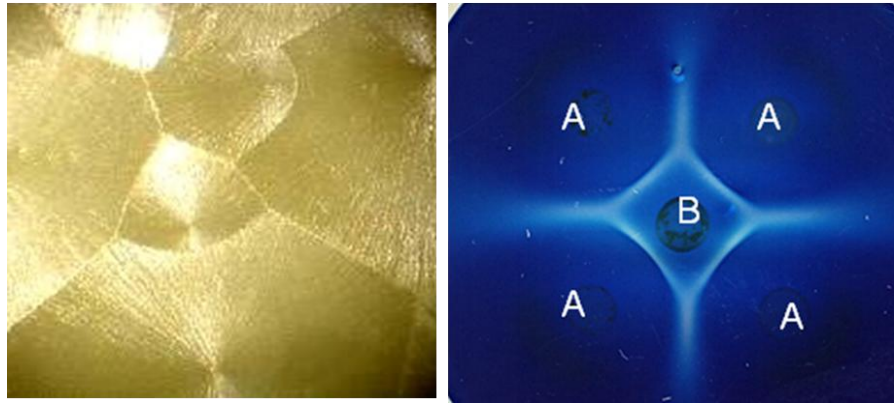

Figure 24 This shows the correct calculation of the bisectors in an addtively weighted Voronoi diagram computed naturally (i.e. without initiation) in growing crystals and b incorrect calculation in a stable chemical processor, the bisectors are have inverted curvature, suggesting an attractive force rather than a repulsive force is responsible for bisector formation.

*4.3 Precipitating Systems vs. Other Unconventional Parallel Processors for Computation of Voronoi Diagrams*

The obvious initial comparison is with the work of Adamatzky on the "hot ice computer" for calculation of a Voronoi diagram [21]. The advantage of the hot ice approach is fast evolution compared to both unstable and stable systems (although all systems could be optimized further). The hot ice computer is reversible and thus it is possible to perform additional (but not sequential) calculations. They all share a permanent output which is advantageous as the results do not have to be reconstructed, and could be used directly to interface with another unconventional processor for shortest path calculations, robot control etc. The advantage of the stable systems over the hot ice computer are the ability to perform sequential calculations, the ability to compute multiple Voronoi diagrams in parallel (upto 3 to date). The ability of the unstable systems over the hot ice approach are the ease of inputs, it is possible to input 100s of points in parallel and because this is done prior to initiating the reaction (simply pouring on the reagent), this is not as difficult as positioning large arrays of input devices. As the hot ice computer is so fast due to the spontaneous nature of the crystallization process then it is difficult to control the evolution of weighted Voronoi diagrams, due to the difficult of inputting two sets of data with a time delay (e.g. for AWVD calculation).

The next obvious comparison is with work undertaken by Adamatzky and de lacy Costello utilizing the Belousov Zhabotinsky reaction to compute a Voronoi diagram [29]. This approach involved controlling the excitability of the BZ reaction to limit natural wave evolution and then initiation of single circular waves using arrays of silver wires. In this respect it shares similarities with the approach used for initiation of the hot ice computer, and also the approach for making controllable waves in unstable precipitating reactions i.e. limiting the natural reactivity to obtain a "controllable "substrate. How does this approach compare? It is still slower than the hot ice computer but has the ability to be optimized. It is marginally faster than the precipitating reactions if reacted within the same area with same input density. It does not give a permanent output and so image analysis is required to reconstruct the Voronoi diagram calculated by the interacting wave fronts (this is because wave fronts in BZ reaction must annihilate). It is inherently difficult to control the inputs beyond a certain number, as with the hot ice computer (but this may simply be an engineering challenge rather than a computational limitation).

As mentioned Voronoi diagrams can be calculated in gas discharge systems [12, 30]. The computation is fast and reversible, but data input can be problematic and the approach is relatively expensive compared to precipitation methods. Also in gas discharge systems it is not apparent how certain weighted diagrams could be computed or indeed multiple diagrams in parallel.

Voronoi diagrams and skeletonisation has also been undertaken using Physarum polycephalum [31], bacterial and fungal cultures [9, 10]. These systems are probably slower than the computation in precipitation systems (although again this could be optimized). They can only compute one Voronoi diagram in parallel and results are not reversible although the computation can be altered during calculation which is a useful function. The bisectors are not formed in the same way as the chemical systems in that they do not leave permanent outputs of the calculation (at least not as clearly). The accuracy of calculation is probably lower than the chemical case because growing fronts tend to be discontinuous. They should be able to compute weighted diagrams, for example if smaller amounts of innoculum are added in combination with larger sources then programmable MWCGVD should be created as the distinct sources should grow with different rates. Figure 25 shows an example of an output from an experiment to calculate a MWCGVD using *Physarum polycephalum*. It can be seen that smaller sources of innoculum have resulted in slower growing fronts which has led to curvature in some of the calculated bisectors. However it is unclear whether the bisector curvature is correctly calculated, or indeed the accuracy of the Voronoi diagram calculation overall. Also presumably if sources are added later in the evolution of the overall calculation then it should be possible to calculate an approximation of an additively weighted Voronoi diagram.

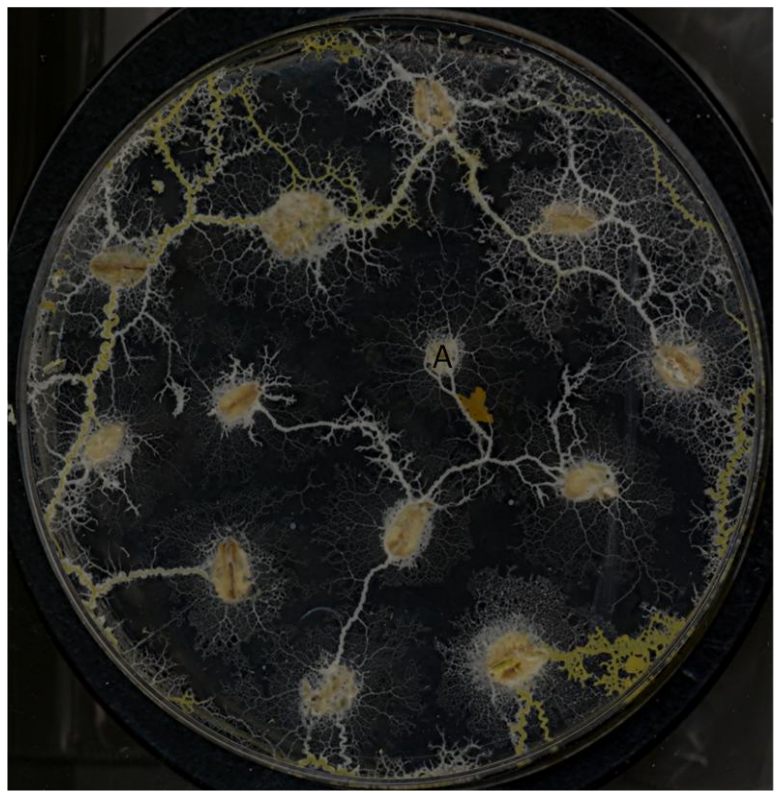

Figure 25 showing attempted calculation of an additively weighted Voronoi diagram using sources of Physarum polycephlum innoculum. The innoculum marked A was half the size of the other innoculum sources. It is apparent that there is some curvature with respect to the source from the growing front at the top right hand side of the image.

In summary there are many unconventional approaches to calculating Voronoi diagrams, all of which have some benefits but also some drawbacks. One thing they all possess is the advantage of parallel computation which is especially beneficial for geometric calculations. However, another thing they all share is slow speed. Additionally they all possess varying degrees of accuracy and some calculations have to be extracted using conventional processing power. However, they are capable of easily solving complex computational problems which some conventional processors would find challenging. However, conventional processors although inefficient for certain geometric calculations possess incredibly high speed and huge numbers of elementary processing units.

## 5. Conclusions

We have shown that Voronoi diagrams can be constructed in "stable" precipitating reactions via the addition of drops of an outer electrolyte to a gel containing an inner electrolyte. The bisectors of the Voronoi diagram are low precipitate areas separating coloured cells. A mechanism of substrate competition between the advancing fronts is suggested as the reason for bisector formation.

A range of complex and colourful tessellations may be obtained if two or more reagents are reacted with the substrate loaded gel. This is particularly true if one or more of the reagents has limited reactivity with the gel. In this case two Voronoi diagrams may be constructed in parallel where the additional diagram corresponds to the original positions of the reactive drops. Even more complex tessellations may be constructed if the products are cross-reactive. This may have some use in materials synthesis as the complex tessellation involves precipitate and precipitate free areas.

If two exclusive chemical couples (ferric ions/ferrocynaide gel and ferrous ions ferricyanide gel) are combined then three Voronoi diagrams may be constructed in parallel. One corresponds to the position of all drops and one corresponds to the position of ferric or ferrous ions.

If the concentration of substrate in the gel is raised then the chemical reactions become unstable and the fronts spontaneously split. Thus Voronoi diagrams are not formed and the controllable pattern formation is lost. If the substrate level is raised even further then the reactions become unstable in three dimensions leading to the formation of growing conical waves. As the front advances through the thickness of the gel these conical waves grow and eventually collide where they annihilate in the formation of a Voronoi diagram. If the concentration is reduced slightly then these conical waves are not formed unless the gel is pre-marked with a glass needle. By using this method Voronoi diagrams could be calculated accurately and fairly rapidly (dependent on heterogenity density). The remarkable feature of these controllable chemical systems is that the higher the density of information the faster the calculation is completed (upto the theoretical maximum).

We tried to implement the same approach with the newly discovered aluminum chloride sodium hydroxide reaction. However, if the gel was marked prior to initiation disorganized waves were generated albeit at specific locations. Therefore, the gel was marked at various time intervals after initiation of the reaction. It was found that wave type could be fairly reproducibly selected. For example if the gel was marked one minute after initiation cardioids waves were the predominant wave initiated. However, if the gel was marked two minutes after initiation then circular waves predominate. Therefore, we were able to construct a range of simple Voronoi diagrams. We also found that marking the gel repeatedly art different time intervals after initiation resulted in the selective initiation of circular waves. Therefore, target like waves, weighted Voronoi diagrams and other user define patterns could be constructed. This work could be very useful in the synthesis of functional materials. This would particularly be true if the effect was reproduced in a number of other inorganic and polymerization type reactions. Also a better method of control rather than physical marking of the gel would be desirable.

This paper has highlighted the wide range of precipitating chemical reactions which are capable of geometric calculations and thus can be classed as specialized chemical parallel processors. It would be desirable to gain a better understanding of the underlying chemical and physical mechanism responsible for the pattern formation in terms of exerting better control over the reactions especially at smaller scales. If this were possible this type of wave interaction computing could be used to design functional materials. Better methods of data input and truly reversible adaptable systems could also be realized.